\documentclass{elsart}
\usepackage{graphicx}
\usepackage{amsmath,amssymb}

\providecommand{\norm}[1]{\lVert#1\rVert}

\begin{document}

\begin{frontmatter}
\title{Track fitting in slightly inhomogeneous magnetic fields}
\author[Madrid]{J.~Alcaraz~\thanksref{CICYT}}
\address[Madrid]{CIEMAT, Avda. Complutense, 22, 28040-Madrid, SPAIN}
\thanks[CICYT]{Partially supported by CICYT Grant: ESP2003-01111}
\begin{abstract}
  A fitting method to reconstruct the momentum and 
direction of charged particles in slightly inhomogeneous magnetic fields 
is presented in detail. For magnetic fields of the order of $1~\mathrm{T}$ 
and inhomogeneity gradients as large as $1~\mathrm{T/m}$ the typical momentum
bias due to the proposed approximations is of the order of few MeV, 
to be compared 
with scattering components of the order of $20$ MeV or even larger.
This method is currently being employed in the reconstruction programs
of the AMS experiment.
\end{abstract}
\end{frontmatter}

\section{Introduction}

   The next generation of particle physics 
experiments will be characterized by an unprecedented accuracy in the 
determination of the positions and momenta of very energetic charged 
particles. The principle of momentum measurement is, in all cases, the 
linear relation between 
the track curvature and the inverse of the momentum in a plane perpendicular 
to the magnetic field direction. High energy physics experiments 
have successfully employed this principle for many decades, 
producing many relevant physics results and discoveries. 

    Track fitting in inhomogeneous magnetic fields
involves the propagation of track parameters between consecutive detector 
layers. Typical approaches~\cite{geane,usual_tracking}, use 
numerical methods of high order in order to integrate the 
equations of motion from layer to layer. This report describes a simple
alternative algorithm, currently being employed in the AMS
experiment~\cite{AMS}. 
All propagation operations are expressed in terms of path integrals, which
are approximated with enough accuracy at initialization. The fitting step is 
also reduced to a linear problem. 
The simplicity of the algorithm allows fast refitting of tracks, even
beyond the reconstruction phase, i.e. at the level of final physics analyses.

   The report is organized as follows.
   The approximations that are the basis of the method and few formulae
to estimate its accuracy are presented in section~\ref{sec:basis}.
The track fitting logic and a typical implementation are described in 
section~\ref{sec:fitting}. Section~\ref{sec:ms} discusses the inclusion of 
multiple scattering effects.
The report is summarized in section~\ref{sec:summary}.

\section{Basis of the method \label{sec:basis}}

    The trajectory of a particle with charge $q$ in a static magnetic field 
$\vec{B}$ is governed by the equation:
\begin{eqnarray}
  \frac{d\vec{p}}{dt} & = & q \left (\vec{v} \times \vec{B} \right )
      \label{eq:main00}
\end{eqnarray}
where $\vec{p}$ and $\vec{v}$ are the momentum and velocity of the particle 
at a given position $\vec{x}$ and time $t$ on the trajectory. 
An immediate conclusion is that
$v\equiv \norm{\vec{v}}$ and $p\equiv \norm{\vec{p}}$ are 
constants of motion. Locally, the trajectory is a helix, with 
$p\sin\theta_B = q \norm{\vec{B}} R$, where 
$R$ is the (signed) radius of curvature in a plane transverse to $\vec{B}$ and 
$\theta_B$ the angle between $\vec{v}$ and $\vec{B}$. Equation~\ref{eq:main00} 
can be rewritten in a different way:
\begin{eqnarray}
\label{eq:main0}
  \frac{d\vec{u}}{dl} & = & \frac{q}{p} \left (\vec{u} \times \vec{B}\right ) \\
    \vec{u} & \equiv & \frac{d\vec{x}}{dl} = \frac{\vec{v}}{v} \nonumber
\end{eqnarray}
where $dl\equiv v~dt$ is the differential length traversed by 
the particle. More visually, $\vec{u}$ is a unitary vector tangent 
to the trajectory at the point $\vec{x}$. Integration of this expression 
between two consecutive layers of a tracker detector, denoted by the 
subscripts $0$ and $1$, leads to:
\begin{eqnarray}
  \vec{u}_1 & = & \vec{u}_0 + \frac{q}{p} \int_{l_0}^{l_1} 
dl^\prime \left (\vec{u} \times \vec{B}\right )(l^\prime)
\label{eq:main}
\end{eqnarray}

  The first approximation in our method consists in computing the previous 
equation as follows:
\begin{eqnarray}
  \vec{u}_1 & \approx & 
\vec{u}_0 + \frac{q}{p} 
\int_{0}^{1} d\alpha ~
\left( \vec{x}_1 - \vec{x}_0 \right) \times
\vec{B}\left(\vec{x}_0+\alpha[\vec{x}_1 - \vec{x}_0]\right)
\label{eq:main1}
\end{eqnarray}

that is, computing the integral along the straight line 
connecting $\vec{x}_0$ and $\vec{x}_1$.
The approximation is exact in two cases: a) when the magnetic field 
is homogeneous, and b) in the infinite momentum limit. For the 
homogeneous field case, $\vec{B} (l) \equiv \vec{B}_0$, and
Equation \ref{eq:main} becomes:
\begin{eqnarray}
  \vec{u}_1 & = & \vec{u}_0 + \frac{q}{p} \left( \int_{l_0}^{l_1} 
dl^\prime \vec{u}(l^\prime) \right ) \times \vec{B}_0 = 
\vec{u}_0 + \frac{q}{p} \left( \vec{x}_1 - \vec{x}_0 \right ) \times \vec{B}_0
\end{eqnarray}
where the last equality is obtained by introducing the definition of $\vec{u}$: 
$d\vec{x} \equiv \vec{u} dl$. The result is identical to the one obtained
using Equation \ref{eq:main1}. In the limit of very high 
momentum, case b), $\vec{u}= \frac{\vec{x}_1 - \vec{x}_0}
{\norm{\vec{x}_1 - \vec{x}_0}}$ up to relative corrections of 
order $1/p$, leading trivially again to Equation \ref{eq:main1}.

In order to estimate the accuracy of the approximation in a general case, 
two additional 
expressions are necessary. First, the expression of the magnetic field 
as a series expansion around $l_{1/2}\equiv(l_0+l_1)/2$: 
\begin{eqnarray}
\vec{B}(l) & = & \vec{B}_0 + \vec{B}^\prime_0~(l-l_{1/2}) + \ldots
\end{eqnarray}
Second, the estimate $\Delta(\vec{u})$ of the difference between the 
true $\vec{u}$ vector and 
$\frac{\vec{x}_1 - \vec{x}_0}{\norm{\vec{x}_1 - \vec{x}_0}}$:
\begin{eqnarray}
\vec{u} & = & \frac{\vec{x}_1 - \vec{x}_0}{\norm{\vec{x}_1 - \vec{x}_0}} + 
\Delta(\vec{u});
\nonumber\\
\Delta(\vec{u}) & = & \frac{q}{p}\int_{l_{1/2}}^{l} dl^\prime
\frac{\vec{x}_1 - \vec{x}_0}{\norm{\vec{x}_1 - \vec{x}_0}} \times \vec{B}_0
 ~+~ \mathcal{O} \left[\left( \frac{q}{p} \right)^2, 
               \frac{\norm{\vec{B}^\prime_0}}{\norm{\vec{B}_0}} \right]
\end{eqnarray}

Introducing the previous expressions in Equation~\ref{eq:main} and comparing
with Equation~\ref{eq:main1} one obtains the following correction at first 
order:
\begin{eqnarray}
  \frac{q}{p} \int_{l_0}^{l_1} dl^\prime \Delta(\vec{u})(l^\prime) \times 
\vec{B}^\prime_0~(l^\prime-l_{1/2}) 
\nonumber \\
= \left(\frac{q}{p}\right)^2 \int_{l_0}^{l_1} dl^\prime 
  \int_{l_{1/2}}^{l^\prime} dl^{\prime\prime} (l^\prime-l_{1/2})
\left[ \frac{\vec{x}_1 - \vec{x}_0}{\norm{\vec{x}_1 - \vec{x}_0}} \times 
\vec{B} \right] \times \vec{B}^\prime_0
\nonumber \\
\approx 
\frac{q}{p}~\frac{(l_1-l_0)^3~\norm{\vec{B}^\prime_0}}{12~R}
\end{eqnarray}
where $R$ 
is the approximate radius of the trajectory 
from $l_0$ to $l_1$. The effect 
has to be compared with the corresponding term in
Equation \ref{eq:main1}, of order 
$\frac{q}{p} \norm{\vec{B}_0} (l_1-l_0)$. 
The difference translates into a relative momentum shift of order:
\begin{eqnarray}
  \frac{\Delta p}{p} \approx \frac{l_1-l_0}{12~R} \times 
\frac{\Delta B}{\norm{\vec{B}_0}}
 \approx \frac{0.3~(l_1-l_0)[{\rm m}]~\Delta B[{\rm T}]}{12~p[{\rm GeV}]}
      \label{eq:accu1}
\end{eqnarray}
where $\Delta B$ is the typical variation of the magnetic field between 
$l_0$ and $l_1$.
For instance, inhomogeneities in the magnetic field of the order of 1 T/m
imply uncertainties of order $\Delta p \approx \pm 1.6$ MeV, 
independent of the absolute value of the rigidity. This has to be compared 
with the typical contributions from multiple scattering in silicon detectors.
For AMS-02~\cite{AMS}, optimized in this respect, the expected 
momentum resolution at the lowest momenta suggests multiple scattering 
effects of order $\Delta p \gtrsim 20$ MeV~\cite{AMS}, safely beyond the 
accuracy of the approximation.

   The second approximation concerns the extrapolation of the position vector 
onto the adjacent plane.
Integrating Equation \ref{eq:main0} twice we obtain:
\begin{eqnarray}
 \vec{x}_1 & = & \vec{x}_0 + \vec{u}_0 (l_1 - l_0)
+ \frac{q}{p} \int_{l_0}^{l_1} dx \int_{0}^{x} dy \left(
   \vec{u} \times \vec{B} \right)(y) \nonumber \\
        & = & \vec{x}_0 + \vec{u}_0 (l_1 - l_0)
+ \frac{q}{p} \int_{l_0}^{l_1} dy (l_1-y) \left(  
   \vec{u} \times \vec{B} \right)(y) \label{eq:ontoplane}
\end{eqnarray}

   The $p\rightarrow\infty$ limit reads:
\begin{eqnarray}
 \vec{x}_1 & = & \vec{x}_0 + \vec{u}_0 \norm{\vec{x}_1 - \vec{x}_0}
+ \frac{q}{p} 
\int_{l_0}^{l_1} dy \left(l_1-y \right) 
\left[ \frac{\vec{x}_1 - \vec{x}_0}{\norm{\vec{x}_1 - \vec{x}_0}} \times 
\vec{B}(y) \right]
\end{eqnarray}

  The previous expression, which is linear in $\vec{x}_0$, 
$\vec{u}_0$ and $q/p$, 
does not coincide in general with the exact solution for the homogeneous 
magnetic field case. Nevertheless, we will prove that it is
precise enough for most cases of interest. Let us define a convenient 
orthonormal reference system by
the unitary vectors $\vec{u}_A$, $\vec{u}_B$ and $\vec{u}_C$:
\begin{eqnarray}
 \vec{u}_A & = & \frac{(\vec{x}_1 - \vec{x}_0) - 
\left[\vec{u}_B (\vec{x}_1 - \vec{x}_0) \right] \vec{u}_B}
{\norm{(\vec{x}_1 - \vec{x}_0)}\sin\theta_B} \\
 \vec{u}_B & \equiv & \frac{\vec{B}}{\norm{\vec{B}}} \\
 \vec{u}_C & = & \frac{(\vec{x}_1 - \vec{x}_0) \times \vec{u}_B}
     {\norm{(\vec{x}_1 - \vec{x}_0)}\sin\theta_B}
\end{eqnarray}
Note that $\vec{u}_B$ is the unitary vector in the direction of the 
magnetic field, and
$\theta_B$ the angle between the vectors $(\vec{x}_1 - \vec{x}_0)$ and 
$\vec{B}$. In terms of these vectors, the trajectory in a homogeneous 
field corresponds to:
\begin{eqnarray}
 \vec{u}(l) & = & 
 \lambda_A(l) \vec{u}_A + \lambda_B(l) \vec{u}_B + \lambda_C(l) \vec{u}_C; \\
     & & \\
  \lambda_A(l) & = & \sin\theta_B
 ~\cos \left[\frac{\sin\theta_B}{R}\left(l-l_{ref}\right)\right] \\
  \lambda_B(l) & = & \cos\theta_B \\
  \lambda_C(l) & = & \sin\theta_B
 ~\sin \left[\frac{\sin\theta_B}{R}\left(l-l_{ref}\right)\right]
\end{eqnarray}
with $\theta_B$ and $l_{ref}$ constants. Let us also write $(l_1-l_0)$ as 
an expansion in powers of $1/R$:
\begin{eqnarray}
 (l_1 - l_0) & = & 
\norm{\vec{x}_1 - \vec{x}_0} +
\frac{\norm{\vec{x}_1 - \vec{x}_0}^3\sin^2\theta_B}{24~R^2} + \ldots
\end{eqnarray}

Equation \ref{eq:ontoplane} for the homogeneous case can be then rewritten as:
\begin{eqnarray}
 \vec{x}_1  & = & \vec{x}_0 + \vec{u}_0 (l_1 - l_0)
\nonumber\\
& & + \frac{q}{p} \int_{l_0}^{l_1} dy (l_1-y) \left(  
    (\lambda_A(y) \vec{u}_A + \lambda_B(y) \vec{u}_B + \lambda_C(y) \vec{u}_C) 
          \times \vec{B}_0 \right) \nonumber \\
& = & \vec{x}_0 + \vec{u}_0 \norm{\vec{x}_1 - \vec{x}_0} + \frac{q}{p} 
\int_{l_0}^{l_1} dy \left(l_1-y \right)
   \frac{\vec{x}_1 - \vec{x}_0}{\norm{\vec{x}_1 - \vec{x}_0}} \times 
~\vec{B}_0 
\nonumber \\
& & 
+ \frac{\norm{\vec{x}_1 - \vec{x}_0}^3\sin^2\theta_B}{24~R^2} 
\left[ \vec{u}_0 - 2~\vec{u}_A \right] 
 ~+~ \ldots
\end{eqnarray}
Since the linear term in $q/p$ gives a contribution of order
$\frac{(l_1-l_0)^2}{2R}$, a naive calculation would suggest a 
relative shift in the momentum of order:
\begin{eqnarray}
  \frac{\Delta p}{p} \approx \frac{l_1-l_0}{12~R}
 \approx \frac{0.3~(l_1-l_0)[{\rm m}]~B[{\rm T}]}{12~p[{\rm GeV}]}
      \label{eq:accu2}
\end{eqnarray}

For a benchmark separation of $20$ cm and a magnetic field
of 1 T, we obtain a maximum possible
shift of $\Delta p \approx \pm 5$ MeV. In practice, the effect is even smaller, 
since the missing
correction affects coordinates in directions less sensitive to bending
($\vec{u}_0$ and $\vec{u}_A$). For most experiments, the fitting procedure is
based on the minimization of a function in which 
position measurements in bending and non-bending directions are almost 
decoupled. In this configuration, the correction above will act 
in quadrature, effectively leading to a momentum shift of order:
\begin{eqnarray}
  \frac{\Delta p}{p} \approx \frac{1}{2}\left( \frac{l_1-l_0}{12~R} \right)^2
 \approx \frac{1}{2}~\left[\frac{0.3~(l_1-l_0)[{\rm m}]~B[{\rm T}]}{12~p[{\rm GeV}]}
         \right]^2
      \label{eq:accu3}
\end{eqnarray}

The quoted shift is negligible, even for large magnetic fields, 
like those of LHC and future linear collider detectors. In fact, this 
conclusion is somehow
equivalent to the one reached in Reference~\cite{karimaki} in the context
of homogeneous magnetic fields. There, only
measurements along the direction 
of the impact parameter with respect to the track at each point 
(i.e. the sensitive ``bending'' direction) are considered. This assumption 
leads naturally to a 
linear problem in terms of the curvature parameter~\cite{karimaki}.

In summary, the following approximations are
considered to be accurate enough for most practical cases:
\begin{eqnarray}
 \vec{u}_{1} & \approx & \vec{u}_{0} 
~+~\frac{q}{p} \norm{\vec{x}_1 - \vec{x}_0}
\int_{0}^{1} d\alpha 
   \frac{\vec{x}_1 - \vec{x}_0}
     {\norm{\vec{x}_1 - \vec{x}_0}} \times
 \vec{B}(\vec{x}_0 + \alpha [\vec{x}_1 - \vec{x}_0]) \label{eq:appr1} \\
 \vec{x}_1 & \approx & \vec{x}_0~+~\vec{u}_0 \norm{\vec{x}_1 - \vec{x}_0}
\nonumber\\
& + & \frac{q}{p} \norm{\vec{x}_1 - \vec{x}_0}^2
\int_{0}^{1} d\alpha \left(1-\alpha \right) 
\frac{\vec{x}_1 - \vec{x}_0}{\norm{\vec{x}_1 - \vec{x}_0}} \times 
\vec{B}(\vec{x}_0 + \alpha [\vec{x}_1 - \vec{x}_0]) \label{eq:appr2}
\end{eqnarray}

\section{Track fitting \label{sec:fitting}}

     For simplicity, it is assumed that all tracker sensitive layers 
are parallel to the $z$ direction and that uncorrelated position 
measurements are 
performed along the $x$ and $y$ directions. The extension to more elaborated
geometrical configurations is straightforward, since only simple rotations
of the predictions are involved. An obvious example is that of a detector 
with a radial 
configuration. To deal with it, it is enough to substitute one of the 
$\chi^2$ terms in the expressions presented later by a sum of
residues along the azimuthal direction.

We consider a scenario in which $z$ coordinates are known with infinite 
precision, so they can be fixed to their nominal values.
The inclusion of an additional $z$-term is, nevertheless, a trivial 
extension to the proposed scheme. Multiple scattering effects will be discussed
in the next section.

We need to determine the 
position of the track at the first plane, $\vec{x}_0\equiv (x_0,y_0,z_0)$, 
the tangent vector at the first plane, 
$\vec{u}_0\equiv (u_{0x},u_{0y},u_{0z})$, and the inverse of the rigidity, 
$q/p$. From Equation \ref{eq:appr2} we obtain, on the second plane:
\begin{eqnarray}
 \vec{x}_1 & \approx & \vec{x}_0 + \vec{u}_0 l_{10} 
+ \frac{q}{p} \vec{\beta}_{10} l_{10}^2
\end{eqnarray}
where the following definitions have been introduced:
\begin{eqnarray}
 l_{j,j-1} & \equiv & \norm{\vec{x}_j - \vec{x}_{j-1}} \\
 \vec{\beta}_{j,j-1} & \equiv & 
\int_{0}^{1} d\alpha \left(1-\alpha \right) 
   \left[\frac{\vec{x}_j - \vec{x}_{j-1}}
    {\norm{\vec{x}_j - \vec{x}_{j-1}}} \times 
 \vec{B}(\vec{x}_{j-1}+\alpha[\vec{x}_j - \vec{x}_{j-1}])\right]
\end{eqnarray}

  The integrals $\vec{\beta}_{j,j-1}$ and the lengths $l_{j,j-1}$ 
are stored in an initialization phase. They are 
determined from the measured positions and the magnetic field values
on the line segment defined by $\vec{x}_{j-1}$ and $\vec{x}_j$.

    At the third layer the extrapolation is given by:
\begin{eqnarray}
 \vec{x}_2 & = & \vec{x}_1 + \vec{u}_1 l_{21} 
+ \frac{q}{p} \vec{\beta}_{21} l_{21}^2 \nonumber \\
           & = & \vec{x}_0 + \vec{u}_0 l_{10} 
+ \frac{q}{p} \vec{\beta}_{10} l_{10}^2 + \vec{u}_0 l_{21}
+ \frac{q}{p} \vec{\gamma}_{10} l_{10} l_{21}
+ \frac{q}{p} \vec{\beta}_{21} l_{21}^2 \nonumber \\
           & = & \vec{x}_0 + \vec{u}_0 \left( l_{10} + l_{21} \right)
+ \frac{q}{p} \left( \vec{\beta}_{10} l_{10}^2 
    + \vec{\beta}_{21} l_{21}^2 + \vec{\gamma}_{10} l_{10} l_{21}
      \right)
\end{eqnarray}
where the following path integral definition has been 
introduced (according to Equation \ref{eq:appr1}):
\begin{eqnarray}
 \vec{\gamma}_{j,j-1} & \equiv & 
\int_{0}^{1} d\alpha 
   \left[\frac{\vec{x}_j - \vec{x}_{j-1}}
     {\norm{\vec{x}_j - \vec{x}_{j-1}}} \times
 \vec{B}(\vec{x}_{j-1}+\alpha[\vec{x}_j - \vec{x}_{j-1}])\right]
\end{eqnarray}

  For many cases of interest (AMS-02 for instance), a Simpson method 
with just a few points is enough to calculate $\vec{\beta}_{j,j-1}$ and 
$\vec{\gamma}_{j,j-1}$ with sufficient accuracy.
  In general, the extrapolation to layer $i$ can be written as:
\begin{eqnarray}
 \vec{x}_i & = & \vec{x}_0 + \vec{u}_0 l_{i0} + \frac{q}{p} \left[
\sum_{k=1}^{i} \left(\vec{\beta}_{k,k-1} l_{k,k-1}^2
+ \vec{\gamma}_{k,k-1} l_{k,k-1} l_{ik} \right ) \right]
\end{eqnarray}
where the lengths $l_{ik}$ must be interpreted as follows:
\begin{eqnarray}
   l_{ik} & = & \sum_{m=k+1}^{i} l_{m,m-1}
\end{eqnarray}

The five parameters $(x_0,y_0,u_{0x},u_{0y},q/p)$ are finally
obtained by minimization of the following chi-square: 
\begin{eqnarray}
 \chi^2 & = & \sum_{i=0}^{N-1} 
  \frac{\left(x_{i,meas}-x_i(x_0,y_0,u_{0x},u_{0y},q/p)\right)^2}{\sigma_x^2} 
  \nonumber \\
        & + & \sum_{i=0}^{N-1} 
  \frac{\left(y_{i,meas}-y_i(x_0,y_0,u_{0x},u_{0y},q/p)\right)^2}{\sigma_y^2}
\end{eqnarray}

where $x_{i,meas}$ and $y_{i,meas}$ are the measured positions
on layer $i$ of the tracker, and $\sigma_x$ and $\sigma_y$ are
the tracker position resolutions in the sensitive directions.
  The minimization leads to a linear equation, which can be 
easily solved via matrix inversion.

   Using an even more simplified notation, the $\chi^2$ can be written in 
a more convenient form:
\begin{eqnarray}
 \chi^2 & = & \sum_{i=0}^{N-1} 
  \frac{\left(x_{i,meas}-\sum_{k=1}^{5}F_{ik} p_k\right)^2}{\sigma_x^2}
            + \sum_{i=0}^{N-1} 
  \frac{\left(y_{i,meas}-\sum_{k=1}^{5}G_{ik} p_k\right)^2}{\sigma_y^2}
\end{eqnarray}
where $p_j; j=1,5$ defines the vector of parameters to be determined,
$\vec{p} \equiv (x_0,y_0,u_{0x},u_{0y},q/p)$. The components of the 
matrices $F_{ij}$ and $G_{ij}$ are:
\begin{eqnarray}
   F_{i1} & = & 1 \nonumber \\
   F_{i2} & = & 0 \nonumber \\
   F_{i3} & = & l_{i0} \nonumber \\
   F_{i4} & = & 0 \nonumber \\
   F_{i5} & = & \sum_{k=1}^{i} \left(\beta^x_{k,k-1} l_{k,k-1}^2
+ \gamma^x_{k,k-1} l_{k,k-1} l_{ik} \right ) 
\end{eqnarray}
and:
\begin{eqnarray}
   G_{i1} & = & 0 \nonumber \\
   G_{i2} & = & 1 \nonumber \\
   G_{i3} & = & 0 \nonumber \\
   G_{i4} & = & l_{i0} \nonumber \\
   G_{i5} & = & \sum_{k=1}^{i} \left(\beta^y_{k,k-1} l_{k,k-1}^2
+ \gamma^y_{k,k-1} l_{k,k-1} l_{ik} \right ) 
\end{eqnarray}
with the upper indices $x$ and $y$ denoting the $x$ and 
$y$ components of the vector integrals $\beta_{k,k-1}$ and $\gamma_{k,k-1}$.

\section{Multiple scattering treatment \label{sec:ms}}

    It will be assumed that the amount of traversed material is reasonably 
small and that the momentum range of interest is such 
that energy losses can be safely neglected. In these conditions, 
    multiple scattering between layers $j-1$ and $j$ is taken 
into account by estimating the additional uncertainty induced on the 
director vector $u_j$. This uncertainty depends on:
a) the amount of traversed material in radiation lengths, 
b) the particle momentum and c) its velocity.

    From layer $j-1$ to layer $j$  a particle is traversing the amount of 
material $X_{j,j-1}$, measured in radiation lengths. The {\it rms} angular 
deviations in the $xz$ and $yz$ projections are equal.
Denoting them by 
by $\Delta_{j,j-1}$, they can be approximately parametrized~\cite{ms1,ms2} 
as follows:
\begin{eqnarray}
  \Delta_{j,j-1} & = & 
          \frac{0.0136}{\beta}~\frac{q}{p[{\rm GeV}]} 
             \sqrt{X_{j,j-1}} \left[1+0.038 \ln (X_{j,j-1}) \right]
\end{eqnarray}
where $\beta$ is the velocity of the particle (in $c$ units) and $p$
its momentum expressed in GeV. Note also that the $X_{j,j-1}$ thicknesses 
hide a dependence on the director vectors $\vec{u}_j$.
The previous expression, accurate at 
the few percent level in the range $0.003 \lesssim X_{j,j-1} \lesssim 0.01$ 
\cite{ms2}, does not admit a Gaussian 
treatment, in the sense that the expected additive property as a 
function of the amount of material is not satisfied: $\Delta^2(X+Y) \ne 
\Delta^2(X)+\Delta^2(Y)$. 

  It is convenient to work with Gaussian uncertainties in order to keep a 
$\chi^2$ minimization scheme. In the case of a very small amount of 
traversed material a possible approach is to assume 
the previous formula to be correct for the total amount of 
traversed material and then distribute the remaining deviations in a linear
way at any intermediate plane, i.e. such that the {\it rms} deviation is
always proportional to $\sqrt{X}$. If the total amount of material is 
$X_{tot}$, the suggestion implies:
\begin{eqnarray}
  \Delta_{j,j-1} & \simeq & 
          \frac{0.0136~\left[1+0.038 \ln (X_{tot}) \right]}{\beta}
                         ~\frac{q}{p[{\rm GeV}]}~\sqrt{X_{j,j-1}} 
\end{eqnarray}

 The previous estimate is usually consistent with the quoted 
accuracy of Equation \cite{ms2}. For the AMS-02 silicon tracker, it 
overestimates the {\it rms} deviations at the intermediate planes by 
at most $4\%$. 

  From the fitting point of view, multiple scattering just modifies
the directions at the different layers as follows:
\begin{eqnarray}
\vec{u}_j & = & \vec{u}_{j}(NO~MS) + \sum_{k=1}^{j} \vec{\epsilon}_{k,k-1}
\end{eqnarray}
where $\vec{u}_{j}(NO~MS)$ denotes the calculation in the absence 
of multiple scattering and $\vec{\epsilon}_{j,j-1}$ is a deviation that 
follows a Gaussian of mean zero and width $\approx 
(\Delta_{j,j-1},\Delta_{j,j-1},0)$. The $j$ dependence enters through the 
amount of accumulated radiation lengths $X_{j,j-1}$ between the exit of layer
$j-1$ and the exit of layer $j$.
  At the level of position measurements the modified trajectories read:
\begin{eqnarray}
 \vec{x}_j & = & \vec{x}_j(NO~MS) + 
      \sum_{k=1}^{j-1} \vec{\epsilon}_{k,k-1}
      \left( \sum_{m=k}^{j-1} l_{m+1,m} \right) \equiv 
        \sum_{k=1}^{j-1} \vec{\epsilon}_{k,k-1} l_{jk}
\end{eqnarray}

   There are two possible options to include these additional sources of 
uncertainty in the $\chi^2$. The first one is to fit all these additional 
parameters ($2 (number~of~planes - 2)$) with additional Gaussian constraints 
according to the expected widths. We
will employ a second option, keeping the same number of fitted parameters, 
but building new covariance matrices according to the Gaussian 
uncertainties $\Delta_{k,k-1}$. In the absence of 
multiple scattering, the covariance matrices for the $x$ and $y$ projections
, $V^0_{ij}$ and $W^0_{ij}$ are given by:
\begin{eqnarray}
 V^0_{ij} =  \left(
 \begin{array}{cccc}
   \sigma_x^2 &      0     & \ldots & 0 \\
        0     & \sigma_x^2 & \ldots & 0 \\
     \vdots   &   \vdots   & \ldots & \vdots \\
        0     &      0     & \ldots & \sigma_x^2 
 \end{array} \right)
  ; & 
 W^0_{ij} =  \left(
 \begin{array}{cccc}
   \sigma_y^2 &      0     & \ldots & 0 \\
        0     & \sigma_y^2 & \ldots & 0 \\
     \vdots   &   \vdots   & \ldots & \vdots \\
        0     &      0     & \ldots & \sigma_y^2 
 \end{array} \right)
\end{eqnarray}

   In the presence of multiple scattering, we need to take into account all 
fully correlated sources via $\vec{\epsilon}_{k}$ terms, leading to the 
matrices:
\begin{eqnarray}
 V_{ij} & = & V^0_{ij} + \sum_{m=1}^{min(i,j)-1} \Delta^2_{m,m-1} l_{im} l_{jm} 
  \nonumber \\
 W_{ij} & = & W^0_{ij} + \sum_{m=1}^{min(i,j)-1} \Delta^2_{m,m-1} l_{im} l_{jm}
\end{eqnarray}
Finally, the $\chi^2$ reads:
\begin{eqnarray}
 \chi^2 & = & \sum_{i,j=0}^{N-1} 
  \left(x_{i,meas}-\sum_{k=1}^{5} F_{ik} p_k\right)~V^{-1}_{ij}~
          \left(x_{j,meas}-\sum_{m=1}^{5} F_{jm} p_m\right) \nonumber \\
        & + & \sum_{i,j=0}^{N-1} 
  \left(y_{i,meas}-\sum_{k=1}^{5} G_{ik} p_k\right)~W^{-1}_{ij}~
          \left(y_{j,meas}-\sum_{m=1}^{5} G_{jm} p_m\right).
\end{eqnarray}
or, in matrix form:
\begin{eqnarray}
 \chi^2 & = & 
  \left(\vec{x}_{meas}- F \vec{p}~\right)^T~V^{-1}~
          \left(\vec{x}_{meas}- F \vec{p}~\right) 
  \nonumber \\
          & + &
  \left(\vec{y}_{meas}- G \vec{p}~\right)^T~W^{-1}~
          \left(\vec{y}_{meas}- G \vec{p}~\right) 
\end{eqnarray}

   Formally, its minimization with respect to $\vec{p}$ leads to the solution:
\begin{eqnarray}
  \vec{p} = \left[ F^T V^{-1} F + G^T W^{-1} G \right]^{-1}
            \left[ F^T V^{-1} \vec{x}_{meas} + G^T W^{-1} \vec{y}_{meas} \right]
\end{eqnarray}

  Even if the $\chi^2$ to be minimized seems formally
linear in the parameters $\vec{p}$, multiple scattering 
introduces a dependence on $q/p$ and $\vec{u}_0$ via the covariance 
matrices $V$ and $W$. 
   A convenient way to solve the problem is to minimize the 
$\chi^2$ following an iterative procedure. In a first step, the $\chi^2$ is 
minimized using the diagonal covariance matrices $V^0_{ij}$ and 
$W^0_{ij}$. The minimization is then iterated several times, using the
$V_{ij}$ and $W_{ij}$ matrices determined from the parameters of the previous
step.
  The iterative procedure is rapidly convergent. It may be stopped either 
after a couple of iterations or when some convergence criteria are reached. 
If computing time is not an issue, a convenient choice is to stop
when the difference in rigidity between two consecutive steps is smaller 
than the accuracy of the method (a few MeV). 

\section{Summary \label{sec:summary}}

    We have presented a simple algorithm for track fitting of high energy
particles traversing slightly inhomogeneous magnetic fields. The method
is based on the prior calculation of a few path integrals which depend 
just on the measured positions and a few values of the magnetic field.
The minimization of a $\chi^2$, which presents a linear dependence on the 
track parameters, leads to a simple solution of the problem. 
Multiple scattering is considered in a straightforward and 
user-controlled way. This is particularly important 
when potential detector resolution problems have to be disentangled 
from trivial material budget effects. Compared to other methods, a few 
simple formulae (\ref{eq:accu1}, \ref{eq:accu2} and \ref{eq:accu3}) 
allow for a fast estimate of the expected momentum uncertainties. 
These formulae use as inputs the 
average value of the magnetic field, the typical size of the field 
inhomogeneities and the distance between measuring layers. The uncertainties
translate into a shift of the measured momentum which,
for most cases of interest in present and future high energy experiments, 
are of the order of a few MeV, well below the uncertainties due to 
multiple scattering. 
The method discussed here is being employed in the context of 
the AMS experiment~\cite{AMS}.
Thanks to its intrinsic simplicity, it is also being
used for fast and reliable track fitting at the latest steps of data
reconstruction and analysis.



\end{document}